\documentclass{article}
\usepackage{epsfig,float}

\title{Study of Giant Pairing Vibrations with neutron-rich nuclei.}

\author{Lorenzo Fortunato \\
Dip. di Fisica  'G.Galilei' and INFN,\\
v. Marzolo 8, 35100 Padova, Italy\\
E-mail: fortunat@pd.infn.it}

\begin{document}

\maketitle

\section*{\it \footnotesize  Talk presented in occasion of the VII School-Seminar on Heavy Ion Physics hosted by the Flerov Laboratory (FLNR/JINR) Dubna, Russia from May 27  to June 2, 2002.}

\section*{Abstract}
We investigate the possible signature of the 
presence of giant pairing states at excitation energy of about 10
MeV via two-particle transfer reactions induced by neutron-rich
weakly-bound projectiles. Performing particle-particle RPA calculations 
on $^{208}$Pb
and BCS+RPA calculations on $^{116}$Sn, we obtain the pairing strength 
distribution for two particles addition and removal modes. Estimates 
of two-particle transfer cross sections can be obtained in the 
framework of the 'macroscopic model'. The weak-binding nature of the 
projectile kinematically favours transitions to high-lying states.
In the case of $(~^6He, ~^4He)$ reaction  we predict a
population of the Giant Pairing Vibration  with cross sections
of the order of a millibarn, dominating over the mismatched 
transition to the ground state.

\section{Pairing field and reaction mechanisms.}
\subsection{Introduction.}
Nuclei in interaction with external fields display a wide variety 
of collective vibrations known as giant resonances, associated with
various degrees of freedom and multipolarities. 
The giant isovector dipole resonance and the giant isoscalar quadrupole 
resonance are the most studied examples in this class of phenomena.
A particular mode, that is associated with vibrations in the number
of particles, has been predicted in the 70's\cite{BrBe} and discussed,
under the name of Giant Pairing Resonance,
in the middle of the 80's in a number of papers\cite{Herz}. 
This phenomenon, despite some early efforts aimed to resolve some broad
bump in the high-lying spectrum in (p,t) reactions\cite{Craw},
is still without any conclusive experimental confirmation.
For a discussion, in particluar in connection with two-particle 
transfer reactions, on many aspects of pairing correlations in nuclei 
we refer to a recent review\cite{OeVi}.

We have studied the problem of collective pairing modes at high 
excitation energy in two neutron transfer reactions with the aim 
to prove the advantage of using unstable beam as a new tool to
enhance the excitation of such modes \cite{Fort}. The main point is that
with standard available beams one is faced with a large energy mismatch
that strongly hinders the excitation of high-lying states and favours the 
transition to the ground state of the final system. Instead the 
'optimum' Q-value condition in the  ($^6$He,$^4$He)
stripping reaction suppresses the ground state 
and should allow the transition to 10-15 MeV energy region.
We have performed particle-particle RPA calculations on lead 
and BCS+RPA on tin, as paradigmatic examples of  normal and  
superfluid systems, evaluating the response to the pairing operator. 
Subsequently the two-neutron transfer form factors have been constructed in 
the framework of the 'macroscopic model'\cite{DaVi} and used in 
DWBA computer codes. We have estimated cross-sections of the order of some 
millibarns, dominating over the mismatched transition to the ground state.
Recently we added similar calculations on other much studied targets to 
give some guide for experimental work.

\subsection{The Giant Pairing Vibrations.}
\begin{figure}[h]
\begin{center}
\epsfxsize=16pc 
\epsfbox{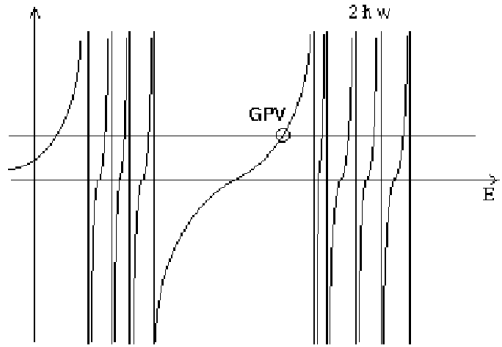}
\end{center} 
\caption{Raw picture of the dispersion relation. The two bunches 
of vertical lines represent the unperturbed energy of a pair of
particles placed in a given single particle energy level.
The graphical solution of the secular equation are the intersection of the 
horizontal line with the curves. The GPV is the collective state relative 
to the second major shell.}
\label{fig1}
\end{figure}

The formal analogy between particle-hole and particle-particle excitations 
is very well established both from the theoretical side\cite{BeBr} and 
from the experimental side for what concern low-lying pairing vibrations 
around closed shell nuclei and pairing rotations in open shells.
The predicted concentration of strength of a $L=0$ character 
in the high-energy region (8-15 MeV for most nuclei) is understood 
microscopically as the coherent superposition of 2p (or 2h) states in the 
next major shell above the Fermi level. 
We have roughly depicted the situation in Fig. (\ref{fig1}). 
In closed shell nuclei the addition of a pair of particles (or holes) 
to the next major shell, with a total energy $2\hbar \omega$, is expected 
to have a high degree of collectivity. Also in the case of open shell 
nuclei the same is expected for the excitation of a pair of particles
with $2\hbar\omega$ energies.

\section{Details of calculations.}
For normal nuclei the hamiltonian with a monopole strength interaction reads:
\begin{equation}
H=\sum_{j} \epsilon_{j} a^{\dagger}_{j}a_{j} - 4\pi G P^{\dagger}P,
\label{p1}
\end{equation}   
where $P$ annihilates a pair of particles coupled to $0$ total angular 
momentum. 
\begin{figure}[b]
\begin{flushleft}
\epsfxsize=14pc
\epsfbox{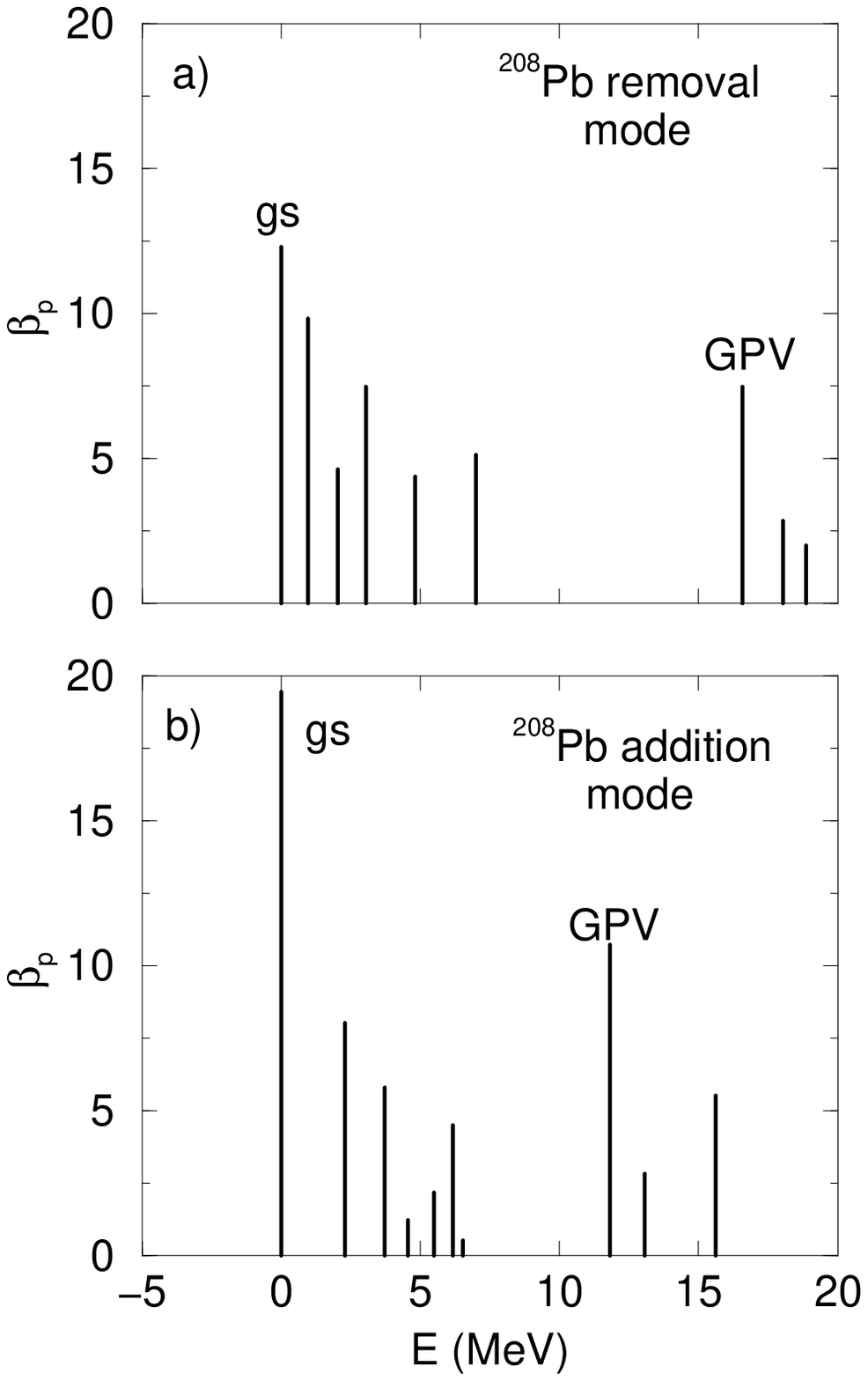}
\end{flushleft}
\begin{flushright}
\vspace{-22.3pc}
\epsfxsize=14pc
\epsfbox{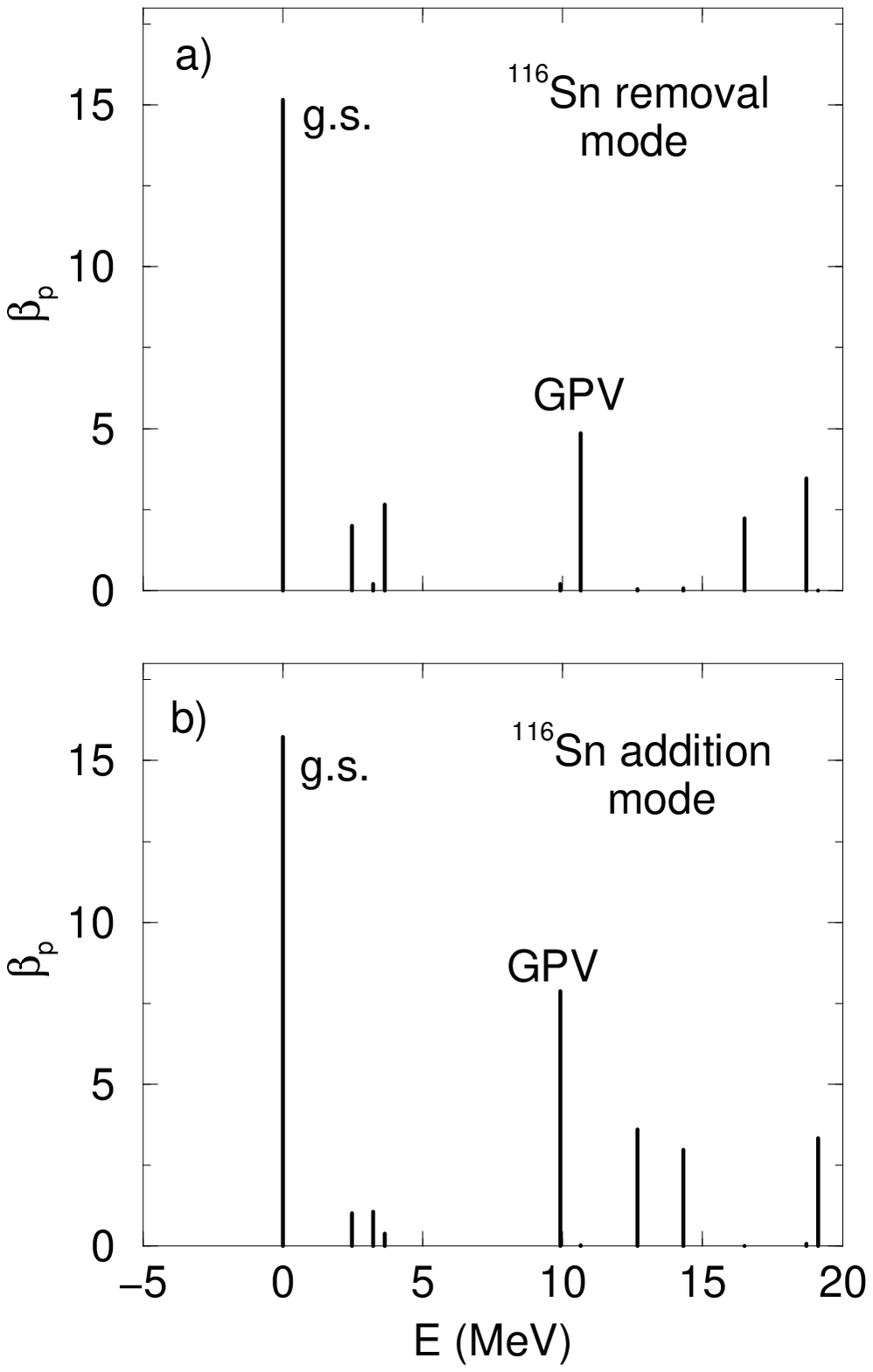}
\end{flushright}
\caption{Pairing response for removal and addition mode in $^{208}$Pb and 
$^{116}$Sn. The ground-state transition and the candidate 
for the GPV are marked.}
\label{fig2}
\end{figure}    
Getting rid of all the technicalities of the solution of the pp-RPA
equations (that may be found in the already cited work by the author) 
we merely state that the pairing phonon may be expressed as a 
superposition of 2p ( or 2h) states with proper forward and 
backward amplitudes ($X_n$ and $Y_n$). The pair transfer 
strength, that is a measure  of the amount of collectivity of a 
each state $n$, is given by:
\begin{equation}
\beta_{Pn} = \sum_{j} \sqrt{2j+1} [X_{n}(j) + Y_{n}(j)] .
\label{p5}
\end{equation} 
This quantity is plotted in the first column of fig. (\ref{fig2}) for the 
removal (upper panel) and addition mode (lower panel). In the same figure
are reported the pairing strength parameters for the states of $^{116}$Sn.
To obtain these last quantities for superfluid spherical nuclei one has to 
rewrite the hamiltonian according to the BCS transformation and has to 
solve more complex RPA equations. In this case the pairing strength for 
the addition of two particles is given, for each state $n$, by:
\begin{equation} 
 \beta_{P}(2p) = \sum_{j} \sqrt{2j+1} \langle n|[a^{\dagger}_{j}
a^{\dagger}_{j}]_{00}|0\rangle = 
\sum_{j} \sqrt{2j+1} [U^{2}_{j} X_{n}(j) + V^{2}_{j}Y_{n}(j)]
\end{equation} 
where the $U$ and $V$ are the usual occupation probabilities. The amount of 
collectivity is a clear signal of the structural existence of giant pairing 
vibrations in the high-lying energy region. We also report here a number of
analogous results for other commonly studied targets
\begin{figure}[b]
\begin{flushleft}
\epsfxsize=9.4pc
\epsfbox{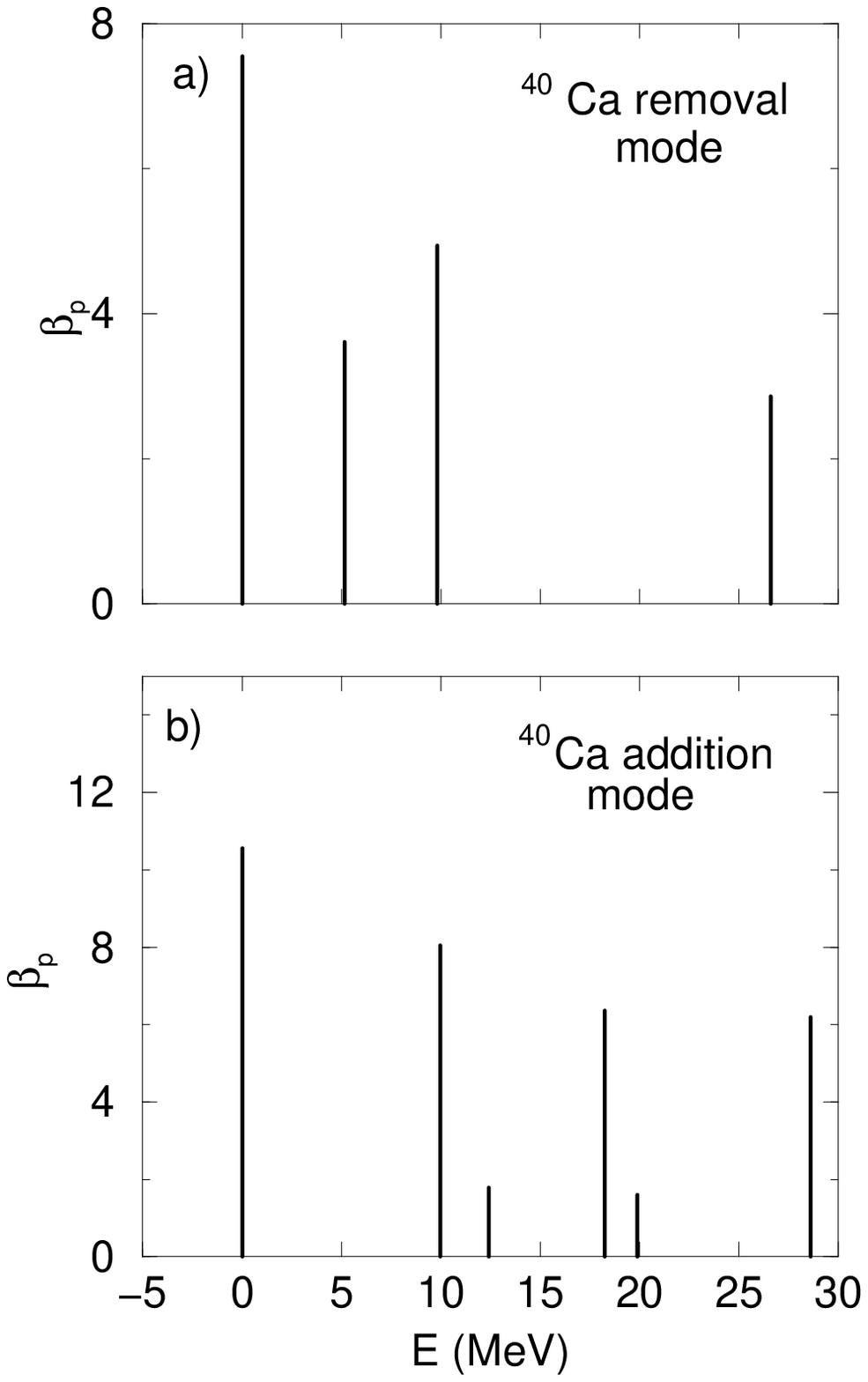}
\end{flushleft}
\begin{center}
\vspace{-15.3pc}
\epsfxsize=9.4pc
\epsfbox{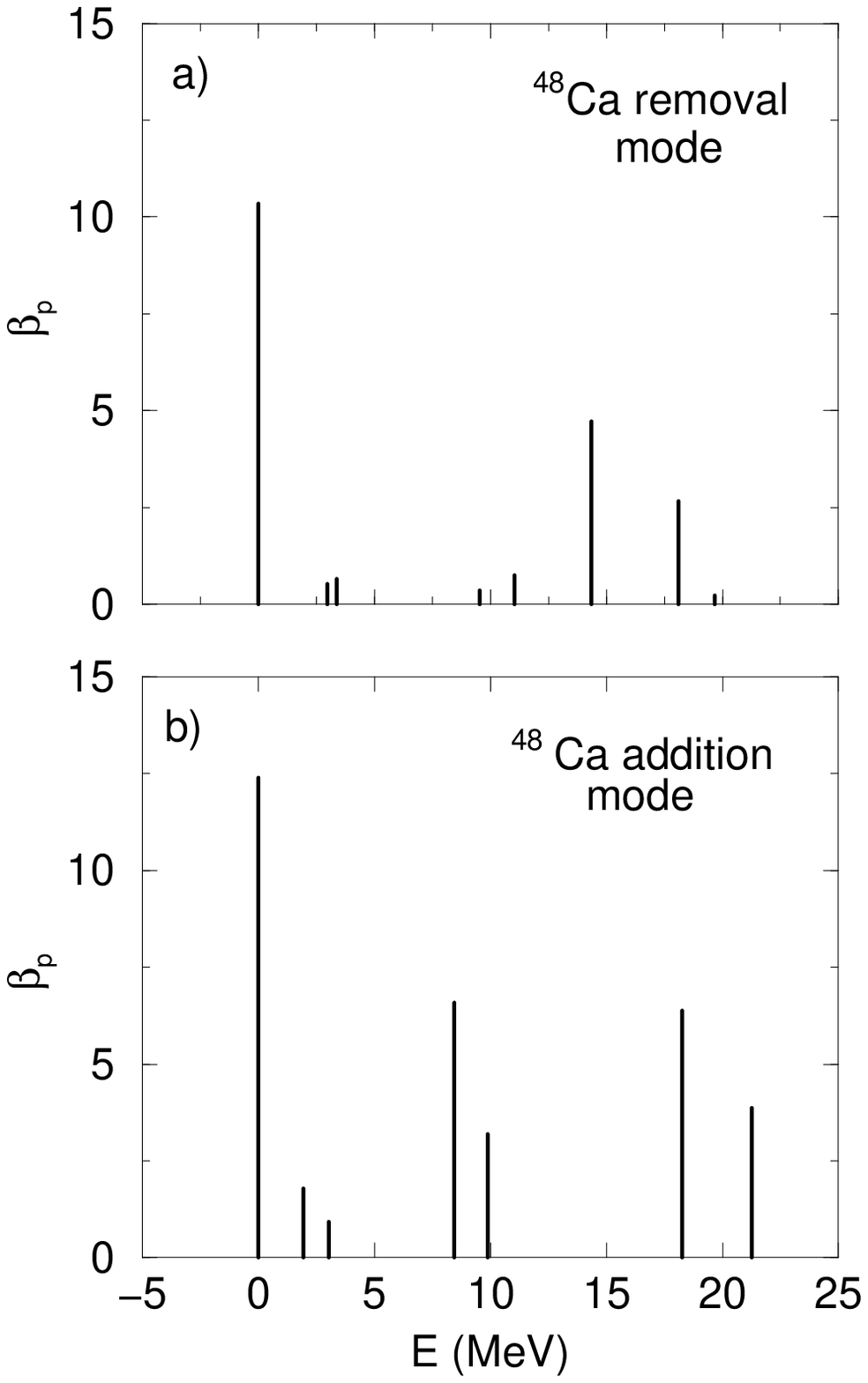}
\end{center}
\begin{flushright}
\vspace{-15.3pc}
\epsfxsize=9.4pc
\epsfbox{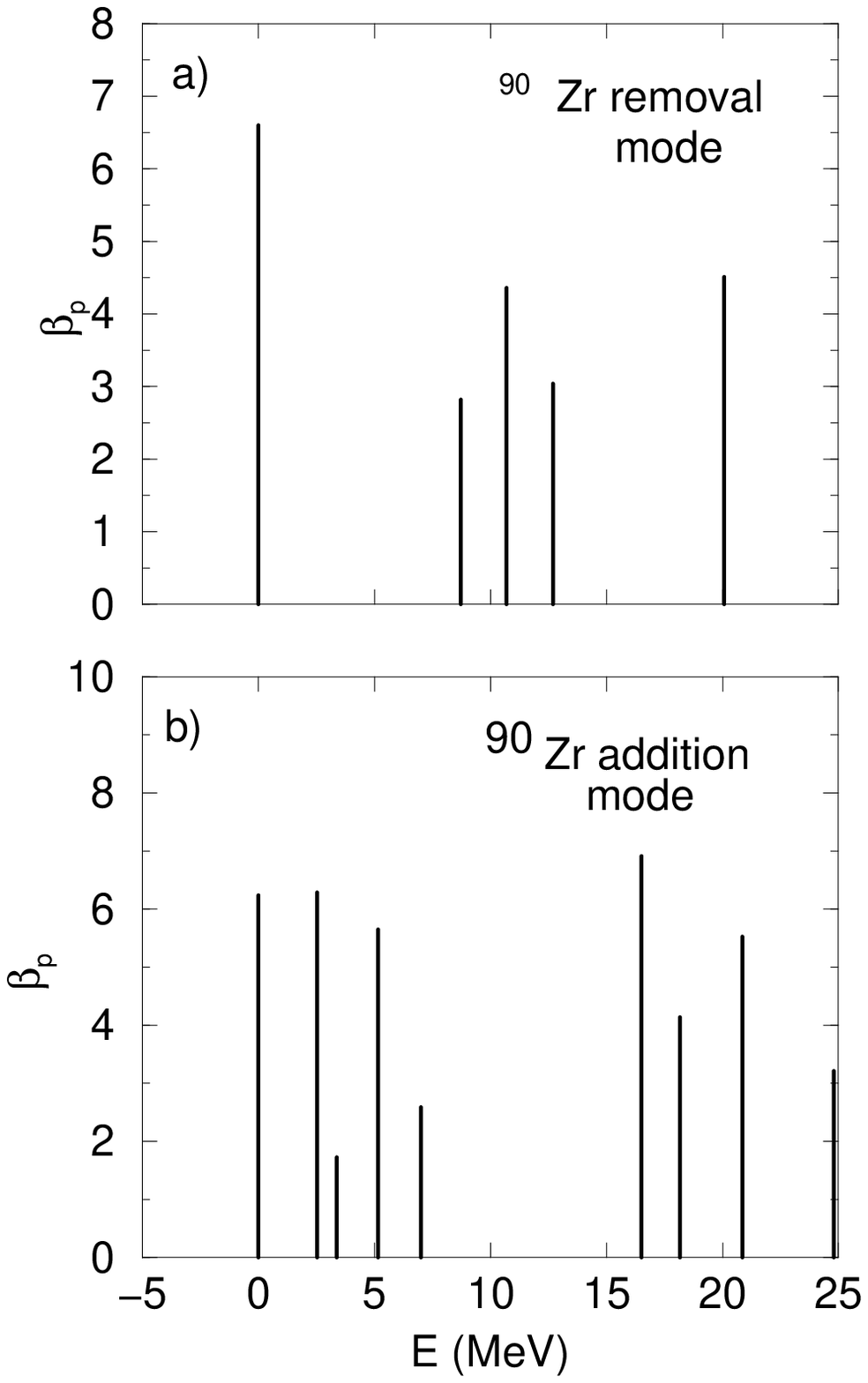}
\end{flushright}
\caption{Pairing response for removal and addition mode in $^{40}$Ca,
 $^{48}$Ca and $^{90}$Zr.}
\label{fig3}
\end{figure}  
with the aim of giving some indications to experimentalists on the reasons 
why we think that lead and tin are some of the most promising candidates.
We have studied two isotopes of calcium with closed shells.
Even if the absolute magnitudes of the $\beta_{\rm P}$ is lower,
it is worthwhile to notice that some enhancement is seen in the more 
neutron-rich $^{48}$Ca with respect to $^{40}$Ca. 
An important role in this change is certainly due to the different shell 
structure of the two nuclei as well as to the scheme that we implemented 
to obtain the set of single particle levels. The latter is responsible
for the collectivity of the removal modes in both Ca isotopes and also 
for the difficulty in finding out a collective state in the addition modes.  
We display also results for $^{90}$Zr where the strength is much more 
fragmented and the identification of the GPV is more difficult.
In the work of Broglia and Bes estimates for the energy of the pairing 
resonance are given as $68/A^{1/3}$ MeV and $72/A^{1/3}$ MeV for normal
and superfluid systems respectively. Our figures follow roughly 
these prescriptions based on simple arguments (and much more grounded
in the case of normal nuclei) as evident from Table \ref{ta1}.
\begin{table}[h]
\begin{center}
\footnotesize
\begin{tabular}{|c|c|c|}
\hline
\raisebox{0pt}[13pt][7pt]{Nucleus} & Our calculation  & Broglia \& Bes estimate \\ 
\hline
Sn & 12.68 MeV & 14.76 MeV \\
Pb & 11.81 MeV & 11.47 MeV \\
\hline
\end{tabular}
\end{center}
\caption{Comparison of position of GPV between our calculation and the Broglia
and Bes estimate.}
\label{ta1}
\end{table}

\section{Macroscopic model for two-particle transfer reactions.}
The starting point of the 'macroscopic model' for two particle transfer 
reactions is to push further the analogy of the vibrations of the nuclear 
surface 
with the 'vibrations' across different mass partitions. If one imagine an 
idealized space in which a discrete coordinate (the number of particles of
the system) labels different sections of the space, it is plausible to give
an interpretation of pairing modes as back and forth oscillations  
in the number of particles. The role of macroscopic variable in this game 
is played by the quantity $\Delta A$, that is the difference in mass from 
the initial mass partition. Exploiting the analogy with inelastic modes
lead us to construct a macroscopic guess for the pairing transition density 
$\delta\rho_p$ modeled on the surface transition density $\delta\rho_s$:
\begin{equation}
\delta \rho_s = {\partial \rho\over \partial \alpha} \alpha= 
 {\partial \rho\over \partial r} R_0 \alpha
\end{equation}
\begin{equation}
\delta \rho_p = {\partial \rho\over \partial \Delta A} \Delta A= 
\Biggl( {R_0 \over 3A}\Biggr)  {\partial \rho\over \partial r} \Delta A
\end{equation}
One usually identifies $\alpha$ with the deformation parameter $\beta_s$,
and the formal analogy suggests the correspondence with a 'pairing 
deformation' parameter $ \beta_s \Leftrightarrow \beta_p /(3A)$.
This scheme implies the assumption that nuclear density is saturated and 
that a change in the number of particles is strictly related to a change of
volume. The two-particle transfer form factors may then be connected to the 
ion-ion potential $U(r)$ as:
\begin{equation}
F_p(r)= \Biggl({\beta_p \over 3A} \Biggr) R_0 {\partial U(r)\over \partial r}
\end{equation} 
This formalism has been applied to many low-energy aspects of 
two-particle transfer reactions\cite{DaPo,DaV2}. Certainly the macroscopic
approach is liable of improvements when one turns to a microscopic
description, but the predictions may be considered robust for giving
order of magnitude evaluations.

\section{Results for Pb, Sn and other targets.}
DWBA calculations have been performed for two-neutron transfer reactions on 
the two cited targets either with usually available beams ($^{14}$C,$^{12}$C)
either with new unstable ones  ($^{6}$He,$^{4}$He). The last reaction has 
been chosen since it has optimal matching conditions: 
the Q-values for the transition to the ground states of both targets strongly
positive, with the consequence of Q-values to the GPV close to the 
optimum Q-value ($Q_{opt} \sim 0$). This should favour the excitation of
the pairing mode, while the situation with carbon beam is reversed, having 
large (and negative) Q-values for the high-lying energy region and 
small Q-values for the low-lying region.
In table \ref{ta2} we report the angle-integrated cross-sections obtained with
standard DWBA computer codes.
\begin{table}[h]
\begin{center}
\footnotesize
\begin{tabular}{|c|c|c|}
\hline
~~&~~\raisebox{0pt}[13pt][7pt]{$^{14}\mbox{C} \rightarrow ~^{12}\mbox{C}$}
~~&~~$^{6}\mbox{He} \rightarrow ^{4}\mbox{He}$\\
\hline
$^{116}\mbox{Sn} \rightarrow ~^{118}\mbox{Sn}_{gs}$~~& 19.4 mb & 0.4 mb\\
$^{208}\mbox{Pb} \rightarrow ~^{210}\mbox{Pb}_{gs}$~~& 15.3 mb & 1.8 mb\\
$^{116}\mbox{Sn}\rightarrow ~^{118}\mbox{Sn}_{GPV}$~~& 0.14 mb & 2.4 mb\\
$^{208}\mbox{Pb} \rightarrow ~^{210}\mbox{Pb}_{GPV}$~~& 0.04 mb & 3.1 mb\\
\hline
\end{tabular}
\end{center}
\caption{Cross-sections for ground-state and GPV transitions obtained with
the DWBA code Ptolemy. The target (column) and projectile (row) are specified.}
\label{ta2}
\end{table}       
\begin{figure}[t]
\begin{flushleft}
\epsfxsize=13.8pc
\epsfbox{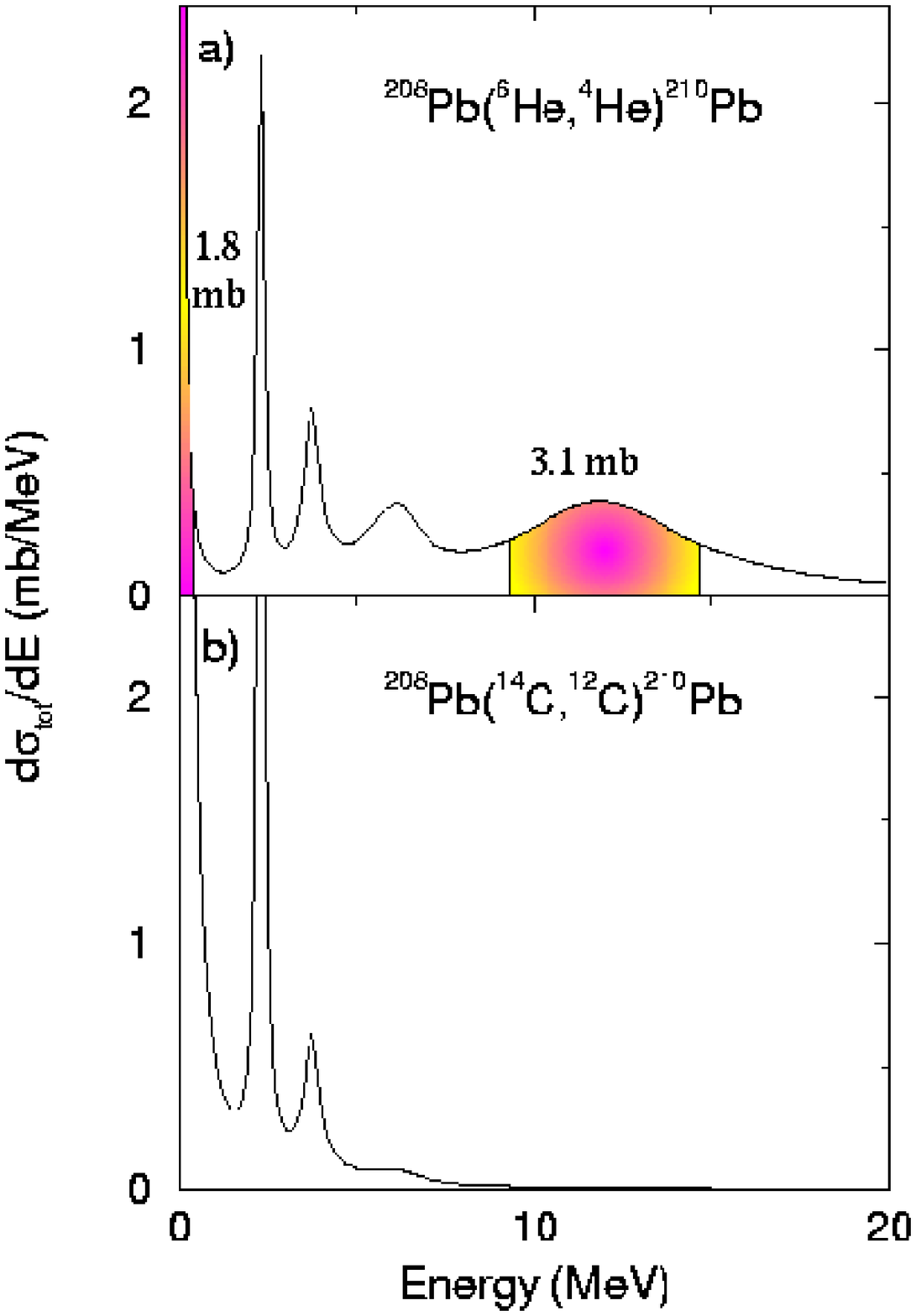}
\end{flushleft}
\begin{flushright}
\vspace{-20pc}
\epsfxsize=13.8pc
\epsfbox{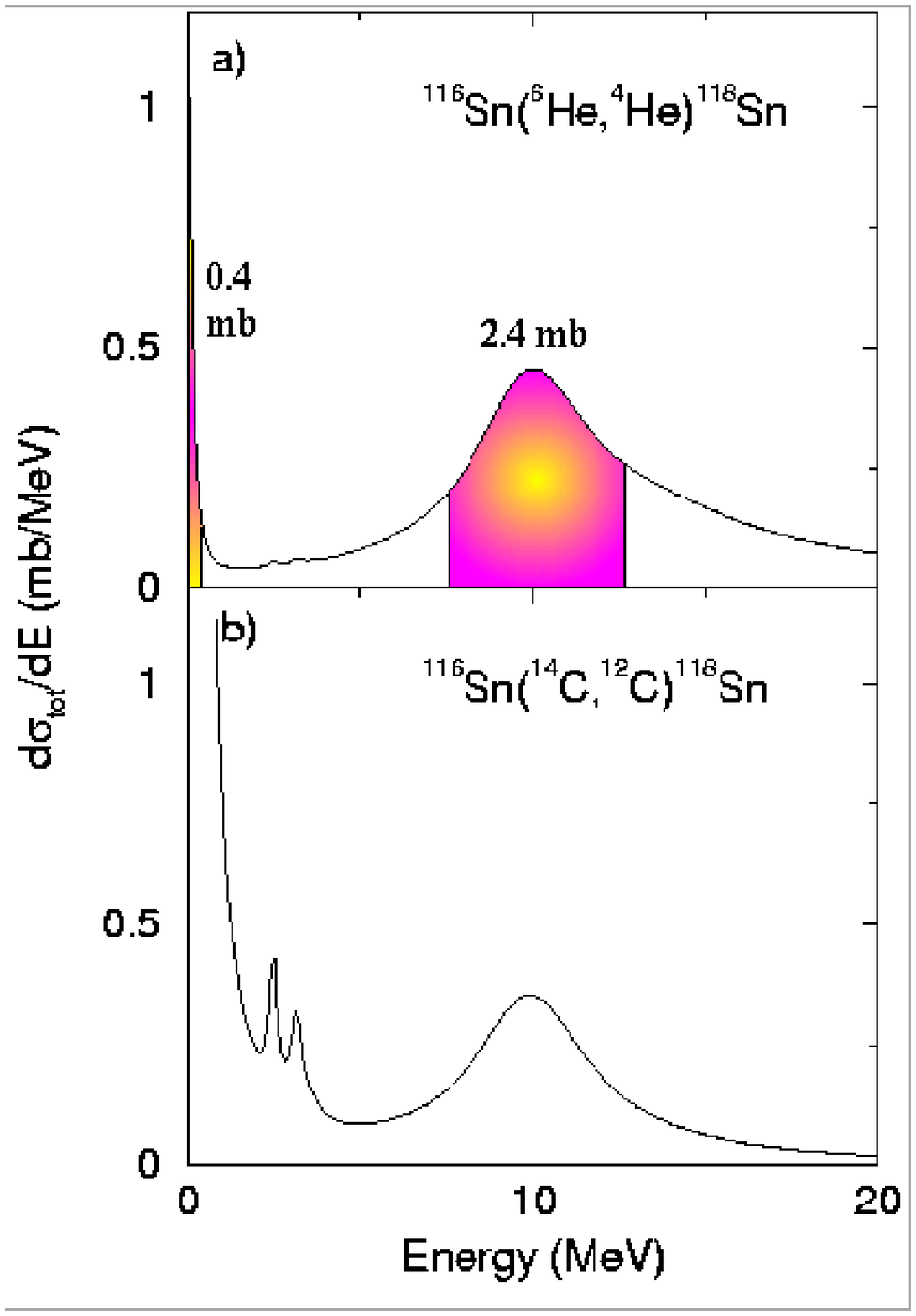}
\end{flushright}
\caption{Differential cross-sections as function of the excitation energy.
The shaded areas for the ($^6$He,$^4$He) reactions allows a comparison between
the transition to the ground states and to the GPV's.
Notice that vertical scale is changed in Sn with respect to Pb.}
\label{fig4}
\end{figure} 
These cross-sections have been derived for sharp states, and we refer
to the numbers in the last table when speaking of order of magnitude estimates.
Obviously cross-section in the high-lying energy region have a finite 
(and large) width that should be inserted for a more realistic description
of the spectrum. We have chosen a simple scheme that gives a lorentzian
distribution with a width that grows quadratically with the excitation energy, 
$\Gamma = k E_x^2$, with $k$ adjusted to give a width of 4 MeV for the GPV.
This could seem rather arbitrary since there is no reason for an {\it a priori}
assignment of this quantity. We have been brought to this simple prescription
because other collective states (of different nature) lying in the same 
energy region display similar values for their width, and it is reasonable to
assume some rule to narrow the low-energy states and to broaden the 
high-energy ones.

\section{Final remarks.}
The final achievements for the four reactions studied in detail are presented 
in Figure \ref{fig4} where the areas corresponding to the cross-sections given
above have been shaded to give a feeling of the relative magnitudes of the 
transition to the ground states and to the GPV's. It is worthwhile 
to note that in the case of Pb there is a considerable gain in using 
unstable beams, while in Sn is much less evident. One sees the need 
for unstable helium when compares the magnitude for the pairing resonance
in the right a) and b) panels with the peak at zero energy: in the first 
panel the transition to the ground state is extremely hindered.

A $^6$He beam is currently available (or it will be available in the very
near future) in many radioactive ion beams facilities around the world and 
the calculations that we have presented could allow a planning for future
experiments aimed to study the not yet completely unraveled role of pairing
interaction in common nuclei, 
using exotic weakly bound nuclei as useful tools. 

\section*{Acknowledgments.}
The author wishes to gratefully acknowledge discussions with Andrea Vitturi,
Hugo Sofia and Wolfram von Oertzen on various aspects of theoretical and 
experimental nuclear physics. The participation at the {\it VII International School-Seminar on Heavy Ion Physics, Dubna, Russia} 2002 has been supported by the INFN.


\begin{thebibliography}{xxxx}
\bibitem{BrBe} R.A.Broglia and D.R.Bes, PLB{69}{129}{1977}.
\bibitem{Herz} M.W.Herzog, R.J.Liotta and L.J.Sibanda, {\it Phys. Rev.} C 
{\bf 31}, 259, (1985).
\bibitem{Craw} G.M.Crawley {\it et al},  PRL {39}{1451}{1977}.
\bibitem{OeVi} W.von Oertzen and A.Vitturi, {\it Rep. Prog. Phys.} {\bf 64}, 
1247-1337, (2001).
\bibitem{Fort} L.Fortunato, W.von Oertzen, H.M.Sofia and A. Vitturi, {\it Eur. J. Phys.} A {\bf 14}, (2002), in press.
\bibitem{DaVi} C.H.Dasso and A.Vitturi (Editors), {\em Collective Aspects in Pair Transfer Phenomena}, SIF Proc. Vol. 18, (Editrice Compositori Bologna, 1987). 
\bibitem{BeBr} D.R.Bes and R.A.Broglia, {\it Phys. Rev.} C {\bf 3}, 2349, (1971).
\bibitem{DaPo} C.H.Dasso and G.Pollarolo, PLB {155}{223}{1985}.
\bibitem{DaV2} C.H.Dasso and A.Vitturi, PRL {59}{634}{1987}.
\end{thebibliography}
\end{document}